\documentclass[journal=nalefd,manuscript=article]{achemso}

\usepackage[version=3]{mhchem} % Formula subscripts using \ce{}

\newcommand*\YO{Y$_2$O$_3$}
\newcommand*\eu{Eu$^{3+}$}
\newcommand*\tr{$^7$F$_0\leftrightarrow ^5$D$_0$}
\DeclareMathOperator{\sinc}{sinc} 
\author{Alexandre Fossati}
\affiliation{Chimie ParisTech, PSL University, CNRS, Institut de Recherche de Chimie Paris, F-75005 Paris, France}
\author{Shuping Liu}
\affiliation{Chimie ParisTech, PSL University, CNRS, Institut de Recherche de Chimie Paris, F-75005 Paris, France}
\alsoaffiliation{Shenzhen Institute for Quantum Science and Engineering, Southern University of Science and Technology, 518055 Shenzhen, China}
\author{Jenny Karlsson}
\affiliation{Chimie ParisTech, PSL University, CNRS, Institut de Recherche de Chimie Paris, F-75005 Paris, France}
\author{Akio Ikesue}
\affiliation{World Laboratory, Mutsuno, Atsuta-ku, Nagoya 456-0023, Japan}
\author{Alexandre Tallaire}
\author{Alban Ferrier}
\affiliation{Chimie ParisTech, PSL University, CNRS, Institut de Recherche de Chimie Paris, F-75005 Paris, France}
\alsoaffiliation {Sorbonne Universit\'e , Facult\'e des Sciences et Ing\'enierie, UFR 933, F-75005 Paris, France}
\author{Diana Serrano}
\author{Philippe Goldner}
\affiliation{Chimie ParisTech, PSL University, CNRS, Institut de Recherche de Chimie Paris, F-75005 Paris, France}
\email{philippe.goldner@chimieparistech.psl.eu}

\title{A Frequency-Multiplexed Coherent Electro-Optic Memory in Rare Earth Doped Nanoparticles}


\begin{document}

\begin{tocentry}
\includegraphics{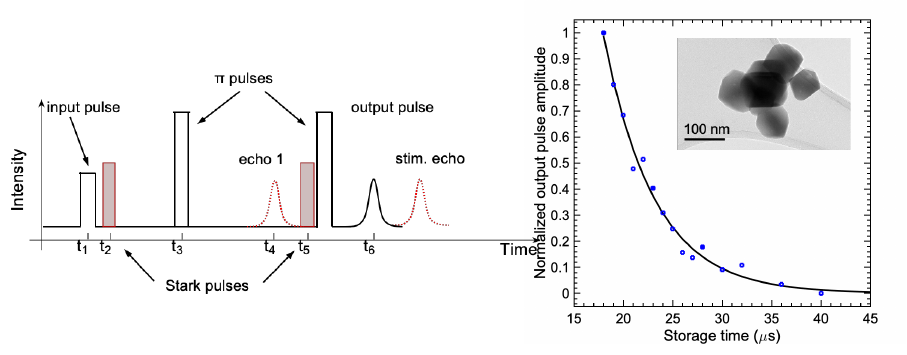}

\end{tocentry}


\begin{abstract}
Quantum memories for light are essential components in  quantum technologies like long-distance quantum communication and distributed quantum computing. Recent studies have shown that long optical and spin coherence lifetimes can be observed in rare earth doped nanoparticles, opening exciting possibilities over bulk materials  e.g. for enhancing coupling to light and other quantum systems, and material design. Here, we report on coherent light storage in \eu:\YO\ nanoparticles using the Stark Echo Modulation Memory (SEMM) quantum protocol. We first measure a nearly constant Stark coefficient of 50 kHz/(V/cm) across a bandwidth of 15 GHz, which is promising for broadband operation. Storage of light using SEMM is then demonstrated for times up to 40 $\mu$s. Pulses with two different frequencies are also stored, confirming frequency-multiplexing capability, and are used to demonstrate the memory high phase fidelity. These results open the way to nanoscale optical quantum memories with increased efficiency, bandwidth and processing capabilities.
\end{abstract}

% \end{abstract}

\section{Main text}

Rare earth (RE) doped crystals are one of the leading solid-state platforms for quantum technologies, offering optical, electron and nuclear spin transitions with long coherence lifetimes $T_2$ at liquid helium temperatures  \cite{Goldner:2015ve,Thiel:2011eu}. In RE doped bulk crystals, optical $T_2$ can reach the ms range \cite{Equall:1994dn,Bottger:2009ik,Businger:2020jf} and nuclear spin state can retain coherence over several hours \cite{Zhong:2015bw}. These materials have been  particularly successful for quantum storage of light \cite{Lvovsky:2009fr,deRiedmatten:2015tj,Awschalom:2018ic} with demonstrations including high efficiency \cite{Hedges:2010dq}, long storage time \cite{Holzapfel:2019ks} and broad bandwidth \cite{Businger:2020jf}. At the nanoscale, stronger coupling between RE and electromagnetic fields or other quantum systems can be exploited. In nanostructured bulk materials for example, single ions and spins can be detected and controlled, opening the way to scalable quantum gates and single photon-spin entanglement \cite{Kindem:2020eb,Raha:2020jb}. Nanomaterials in the form of films and nanoparticles are also attracting a strong interest, as they offer increased flexibility in design, composition and integration \cite{Zhong:2019gf,Scarafagio:2020kl,Singh:2020ey,Dutta:2019fk,Bartholomew:2017ik,Serrano:2018ea,Serrano:2019km}. \eu:\YO\ nanoparticles are especially promising as they can show long optical and spin coherence lifetimes, up to 10 $\mu$s \cite{Liu:2018gk} and 8 ms \cite{Serrano:2018ea} respectively, depending on particle size, synthesis and post-treatment. They can be inserted in tunable fiber micro-cavities to increase light-ion coupling and increase fluorescence rates through the Purcell effect, as demonstrated for \eu\ \cite{Casabone:2018kc} and Er$^{3+}$ ions \cite{Casabone:2020wj}. Although these results are promising, functionalities that  exploit quantum coherence have not yet been reported in these nanomaterials. 

Here, we demonstrate coherent storage of light in 100 nm \eu\YO\ nanoparticles using an on-demand electro-optic quantum protocol. Memory time up to 40 $\mu$s is demonstrated, as well as frequency-multiplexed storage, a key requirement for efficient quantum memories \cite{Lvovsky:2009fr}. The phase fidelity of the memory is investigated and found to be very high, with an input-output phase correlation over 0.99. Moreover, measurements of Stark coefficient indicate that the storage bandwidth could be extended over several GHz. These results suggest that  RE nanoparticles are promising materials for multimode and long storage time nanoscale quantum memories for quantum communications, distributed quantum computing and other applications relying on coherent light-matter interfaces.

% \begin{itemize}
%     \item rare earth materials for QT and nano-systems: bulk \cite{Goldner:2015th,Thiel:2011eu}, bulk and cavity \cite{Zhong:2017fe}\cite{Kindem:2020eb}\cite{Raha:2020jb},general \cite{Zhong:2019gf}, films \cite{Scarafagio:2020kl}\cite{Singh:2020ey}\cite{Dutta:2019fk}, nanos \cite{Bartholomew:2017ik}\cite{Serrano:2018ea} with cavity \cite{Casabone:2020wj,Serrano:2019km}\cite{Casabone:2018kc}
    
%     \item memories \cite{Northup:2014gv} in RE \cite{Goldner:2015th}, \cite{Bussieres:2014dc},\cite{Maring:eb} protocols \cite{deRiedmatten:2008ck}
%     \item current achievements
%     \item here \cite{Arcangeli:2016eu}, \cite{Damon:2011tx},\cite{McAuslan:2011ke},\cite{Liu:2018gk} \cite{Macfarlane:2014fy}
% \end{itemize}

The storage protocol we use is called Stark Echo Modulation Memory (SEMM) \cite{Arcangeli:2016eu}. It uses double resonant rephasing \cite{Damon:2011tx,McAuslan:2011ke,Afzelius:2013ga} and phase control by electric field induced frequency shifts to allow for low-noise and broadband quantum storage. We previously demonstrated SEMM for RE nuclear spins in a bulk crystal, storing and retrieving coherent RF excitations with high fidelity over the entire spin inhomogeneous linewidth \cite{Arcangeli:2016eu}. The work presented here is the first to investigate this protocol in the optical domain.

\begin{figure}
  \includegraphics[width = \textwidth]{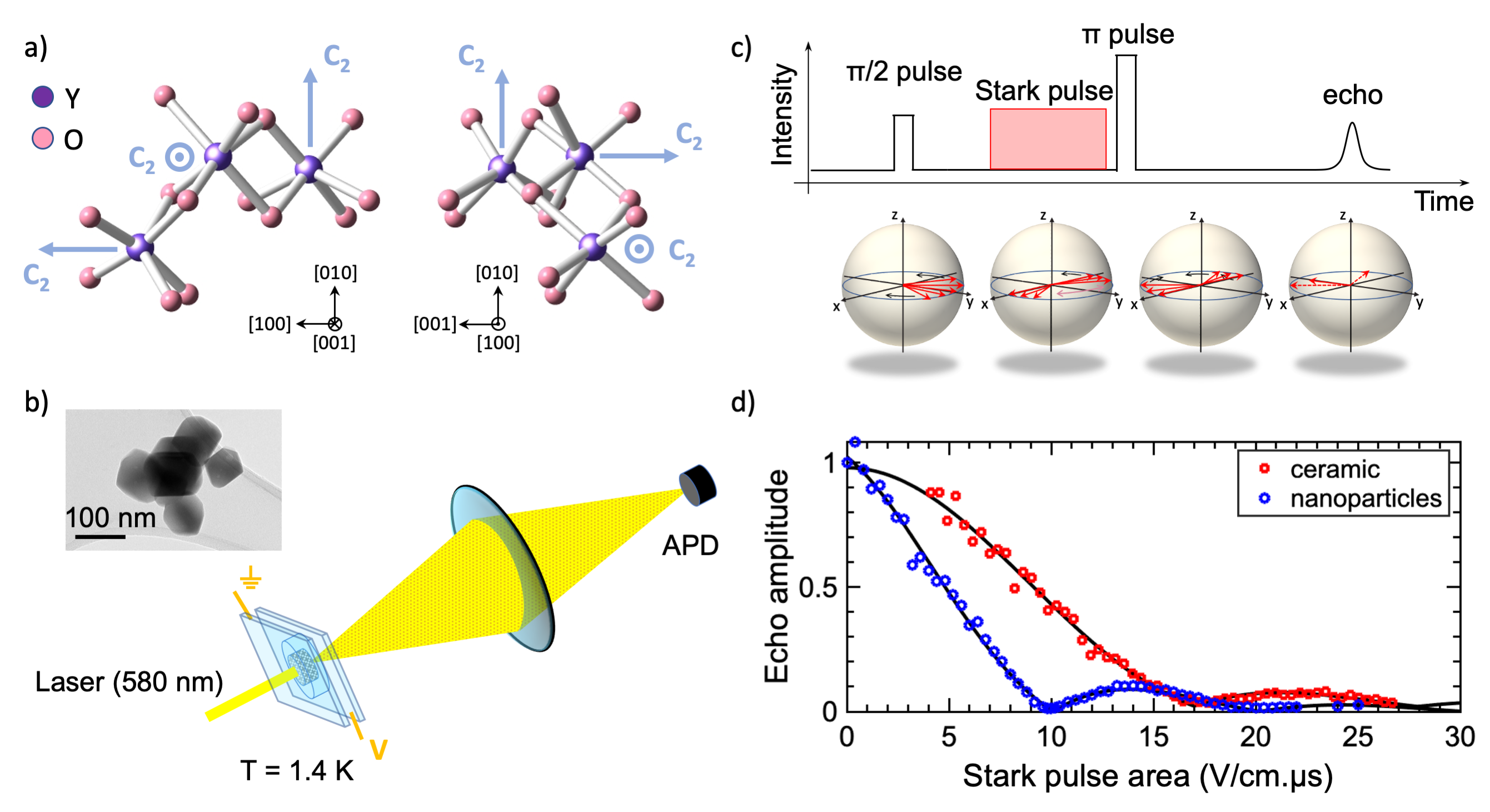}
  \caption{Stark effect measurements in \eu:\YO. (a) Non equivalent Y$^{3+}$  sites (3 out of 6) with $C_2$ symmetry in \YO\ crystal viewed along two different lattice orientations. Permanent and transition electric dipole moments are along the $C_2$ axis. (b) Electron microscope image of 0.3\% \eu:\YO\ nanoparticles and experimental set-up. Nanoparticles in the form of a powder are maintained between two transparent FTO electrodes separated by 0.5 mm and located inside a helium bath cryostat. The laser at 580 nm is focused on the sample and backwards scattered light collected by lenses and directed to an avalanche photo-diode. (c) Top: pulse sequence for the Stark modulated photon echo experiment (black: light, red: electric field). The stark pulse amplitude is varied, whereas other parameters (delays, pulse lengths) are fixed. Bottom: evolution of states in the Bloch sphere. The Stark pulse causes a decrease of the echo amplitude (right-hand sphere). (d) Normalized echo amplitude as a function of the Stark pulse area (see text) for nanoparticles and a transparent ceramic. Solid lines are fits to Eqs. \ref{eqn:echoStarkCeram} and \ref{eqn:echoStarkNano}.}
  \label{fig: fig1}. 
\end{figure}

SEMM is based on centers exhibiting a linear Stark effect, preferably in a  centrosymmetric crystal \cite{Macfarlane:2007it,Kaplyanskii:2002cy}. This is the case for \eu\ ions in the $C_2$ sites of \YO\ in the cubic phase. The permanent  electric dipole moment $\boldsymbol{\mu}$ of \eu\ levels is oriented along the $C_2$ axis, which is parallel to the $<100>$ family of crystallographic axes (Fig. \ref{fig: fig1}a) \cite{Graf:1997kn}. There are therefore 6 possible orientations for the $C_2$ sites, that can be grouped in three sets of two sites related by an inversion symmetry. \eu\ electric dipole transitions, such as the one of interest here \tr, follow the same rule and are polarized  along the $C_2$ axes. The Stark shift for a transition between ground ($|g>$) and excited ($|e>$) states is proportional to the applied electric field  amplitude $E$ \cite{Macfarlane:2007it}:
\begin{equation}
    \delta \nu =  L \boldsymbol{\delta \mu}\cdot {\bf E} = k E \cos(\theta),
    \label{eqn:StarkShift}
\end{equation}
where $L$ is the Lorentz correction for the material polarizability, $\boldsymbol{\delta \mu} = \boldsymbol{\mu}_e - \boldsymbol{\mu}_g$ the difference between ground and excited permanent dipole moments, $k$ the Stark coefficient and $\theta$ the angle between $\boldsymbol{\delta \mu}$ and $\bf{E}$. The dipole moments are proportional to the matrix elements $<l| \mathbf{r}|l>$, $l = g,e$. Ions in sites with different $C_2$ axis orientations will experience different Stark shifts for an arbitrary electric field and are therefore generally non-equivalent. Those related by an inversion symmetry experience an opposite Stark shift, giving rise to the so-called pseudo-Stark splitting for non-degenerate levels like $^7$F$_0$ and $^5$D$_0$. 

\begin{figure}
  \includegraphics[width = \textwidth/2]{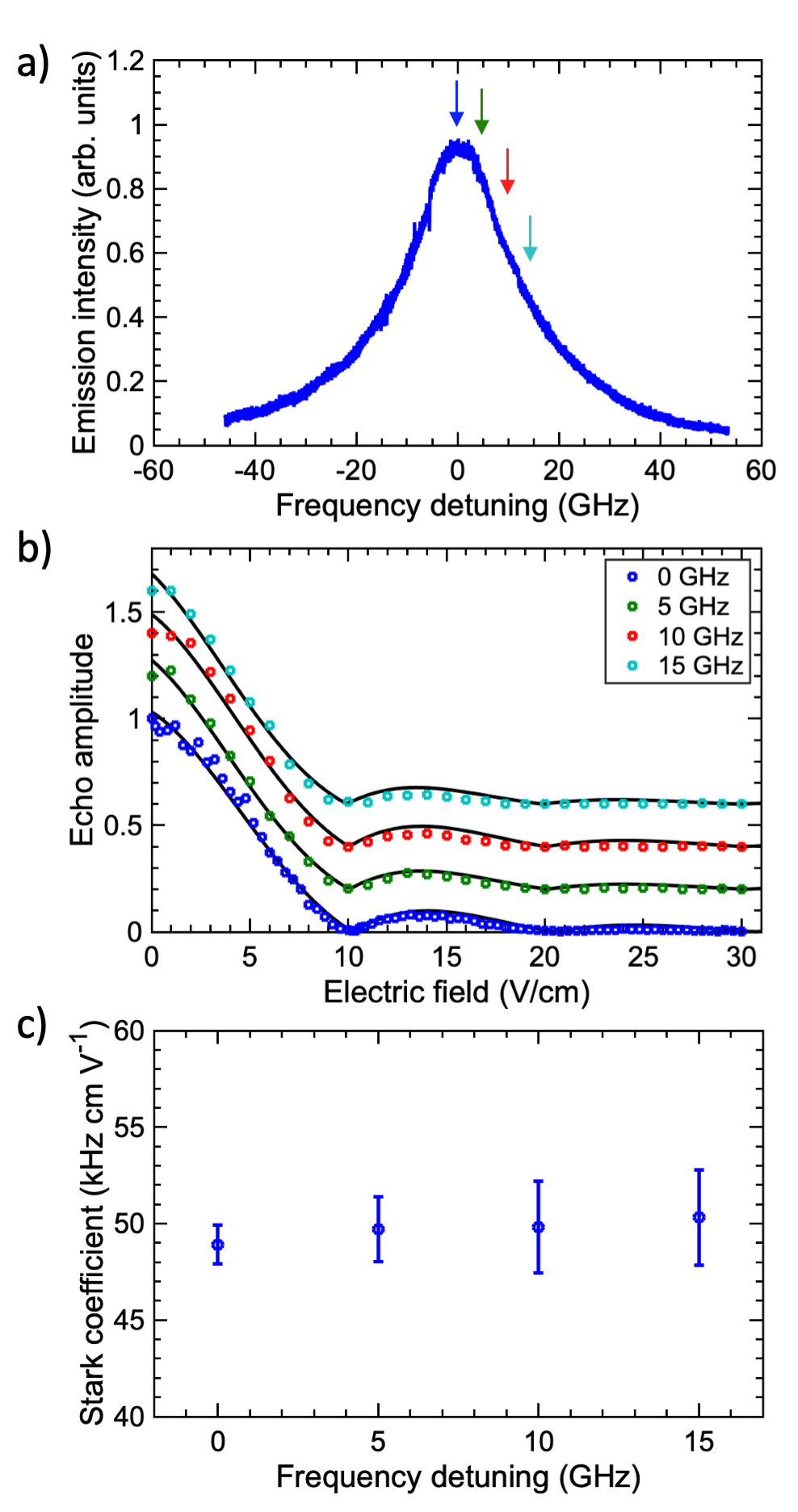}
  \caption{Optical frequency dependence of Stark coefficient. (a) Excitation spectrum of \eu:\YO\ $^7$F$_0 \to ^5$D$_0$  transition  centered at 580.88 nm (vac.).  (b) Normalized echo amplitude as a function of electric field at different excitation frequencies (arrows in (a)). Data are vertically translated for clarity. Solid lines are fits to Eq. \ref{eqn:echoStarkNano}. (c) Fitted Stark coefficient as a function of optical frequency from line center. }
  \label{fig: fig2}
\end{figure}

Measurements were carried out on nanoparticles obtained by homogeneous precipitation followed by high temperature annealing and micro-wave oxygen plasma treatment (see SI for details) \cite{Liu:2020ty}. The non-agglomerated single crystalline  particles had a 100 nm average diameter (Fig. \ref{fig: fig1}b). The experimental set-up is shown in Fig. \ref{fig: fig2}b) and was previously developed for observing photon echoes in strongly scattering media \cite{Perrot:2013hy}. A coherence lifetime of $5.7 \pm 0.2$ $\mu$s at 1.4 K was previously measured on these nanoparticles \cite{Liu:2020ty}. 

We next  used the very sensitive technique of Stark modulated echoes \cite{Mims:1964cj,Meixner:1992uu,Macfarlane:2014fy} to determine \eu\ Stark coefficient $k$. The principle is shown in Fig. \ref{fig: fig1}c: a regular photon echo sequence is set-up and an electric field pulse applied between the $\pi/2$ and $\pi$ pulses. The field induces frequency shifts, corresponding to a rotation along $z$ in the Bloch sphere (Fig. \ref{fig: fig1}c). Ions related by the inversion undergo opposite frequency shift and therefore rotate in opposite direction in the Bloch sphere. Because of this rotation, refocusing after the $\pi$ pulse can result in a lower echo amplitude (Fig. \ref{fig: fig1}c). Specifically, for ions with a $C_2$ axis oriented parallel to $\bf E$, the echo amplitude $A$ follows an oscillatory behavior:
\begin{equation}
    |A| = A_0|\cos(2 \pi k E T_s)|
\end{equation}
where $T_s$ is the duration of the Stark pulse, assumed to have  a square shape.  

Fig. \ref{fig: fig1}d displays the echo amplitude, recorded using heterodyne detection, on the nanoparticles as a function of the Stark pulse area $A_S = E T_s$ (see SI for  details). We did observe an oscillation, with the echo decreasing to zero amplitude for $A_S = 10$ V/cm$\cdot\mu$s and reaching a second maximum at $A_S = $ 14 V/cm$\cdot\mu$s.  The signal is also strongly damped with increasing $A_S$. Such damping can be related to a distribution of Stark coefficients that produces an inhomogeneous broadening not refocused in the echo sequence \cite{Mims:1964cj}. This broadening increases with increasing $A_S$, resulting in a damping. The different site orientations in \YO\ can also lead to echo amplitude modulations \cite{Graf:1997kn}. In our case, however, the Stark coefficient distribution is dominated by the random orientations of the particles, and therefore of the \eu\ $C_2$ axis, with respect to the electric field. Another phenomenon to take into account is the dependence of the echo amplitude on the angle between light polarization and the transition dipole moment (see SI), also oriented along the $C_2$ axis \cite{GorllerWalrand:1996bt}. 

These combined effects are clearly seen when the experiment is performed on a transparent polycrystalline \eu:\YO\ ceramic \cite{Kunkel:2016fh}. In this case, damped oscillations are also observed (Fig. \ref{fig: fig1}d) and can be accurately modeled. Since the electric field and light polarization were set perpendicular, the echo amplitude is expressed as (see SI):
\begin{equation}
    |A| = A_0\int_0^{\pi/2} \cos(2 \pi k A_S \cos\theta)(\sin \theta)^4 d\theta
    \label{eqn:echoStarkCeram}
\end{equation}
Experimental data can be very well fitted by Eq. \ref{eqn:echoStarkCeram} (Fig. \ref{fig: fig1}d), giving  $k =  48.0 \pm 1$ kHz/(V/cm). The nanoparticles behave quite differently (Fig. \ref{fig: fig1}d), which we attribute first to randomization of the light polarization direction by the strong scattering in the powder (see SI). Such a random distribution alone is however unable to reproduce the experimental data and it was found that an additional exponential damping term was needed. We relate it to electric field inhomogeneity due to the large sample volume in which the echo forms because of light scattering and \YO\ high dielectric constant ($\epsilon_r = $ 15 \cite{Robertson:2004fa}). 

% Partially randomized light polarization could also play a role, but distributions of $k$ related to crystal field variations or local electric fields,  seem unlikely, as shown below.  

Nanoparticles data were fitted to the expression:
\begin{equation}
     |A| = A_0|\sinc(2 k A_S )|e^{-b A_S}.
     \label{eqn:echoStarkNano}
\end{equation}
The fit shown in Fig. \ref{fig: fig1}d gives $k =  49.5 \pm 1$ kHz/(V/cm) and a damping factor $b = 6.4 \pm 0.8 \times 10^{-2}$ cm/V/$\mu$s. The Stark coefficient in the nanoparticles is therefore almost identical to that found in the transparent ceramic. Crystallite size has therefore a low influence on $k$, since it is 1-2 $\mu$m in the ceramic \cite{Kunkel:2016fh} and only 100 nm in the particles. It thus suggests that the polarizability correction factor and dipole matrix elements (Eq. \ref{eqn:StarkShift})  in the nanoparticles are close to bulk values, in agreement with other properties, such has optical line position and width, which are also similar to  bulk ones \cite{Liu:2020ty}. 

The \eu:\YO\ Stark coefficient value is comparable to those measured in \eu:Y$_2$SiO$_5$ (27 - 35 kHz/(V/cm)) \cite{Graf:1997kn,Macfarlane:2014fy} or \eu:YAlO$_3$ (33.5 kHz/(V/cm)) \cite{Graf:1997kn} bulk crystals. In the context of RE quantum gates,  $k$ is a fundamental parameter to design pulse sequences and material. Indeed, two-qubit gates are implemented by excitation of the control ion, which causes a shift in the target ion frequency  \cite{Walther:2015hh}. The large value of $k$ found in \eu:\YO\ allows coupling distant ions, and is therefore favorable to a high density of qubits that can interact. However, it also enhances transition sensitivity to electric noise, which is likely to be a major source of dephasing in the nanoparticles \cite{Bartholomew:2017ik,Serrano:2018ea}. Taking full advantage of large Stark shifts for quantum gates will therefore require samples with low concentration of charged defects in volume or surface.

We next investigated the dependence of the Stark coefficient across the nanoparticles inhomogeneously broadened optical transition. Its half width at half maximum is 27 GHz and we measured Stark modulated echoes for detunings of 0, 5, 10 and 15 GHz from line center (Fig. \ref{fig: fig2}a). Variations of echo amplitude as a function of electric field amplitude are very similar despite the  large detunings, although signals reduce when absorption decreases (Fig. \ref{fig: fig2}b). Accordingly, the fitted Stark coefficients exhibit a low dependence on detuning, likely within a few percent, the increase in error bars being essentially related to signal strength and imperfect fit model. This result shows that ions located in perturbed environments, i.e. away from line center, only slightly differ in $\boldsymbol{\delta \mu}$. This suggests  that Stark coefficient broadening due to crystal field (CF) variations is indeed small. This is consistent with the second-order dependence of permanent dipole moments on CF, through changes in wavefunctions. Small variation of $k$ with detuning also indicates that the optical inhomogeneous broadening is not due to interactions between ions and nearby charged defects. A local electric field would be present, modifying the effective Stark coefficient. Such defects, which are likely to be a major cause of dephasing in the nanoparticles, as mentioned above, are therefore rather randomly dispersed. This is in agreement with the nearly constant optical coherence lifetimes observed across the inhomogeneous absorption line \cite{Liu:2020ty}.

\begin{figure}
  \includegraphics[width = \textwidth]{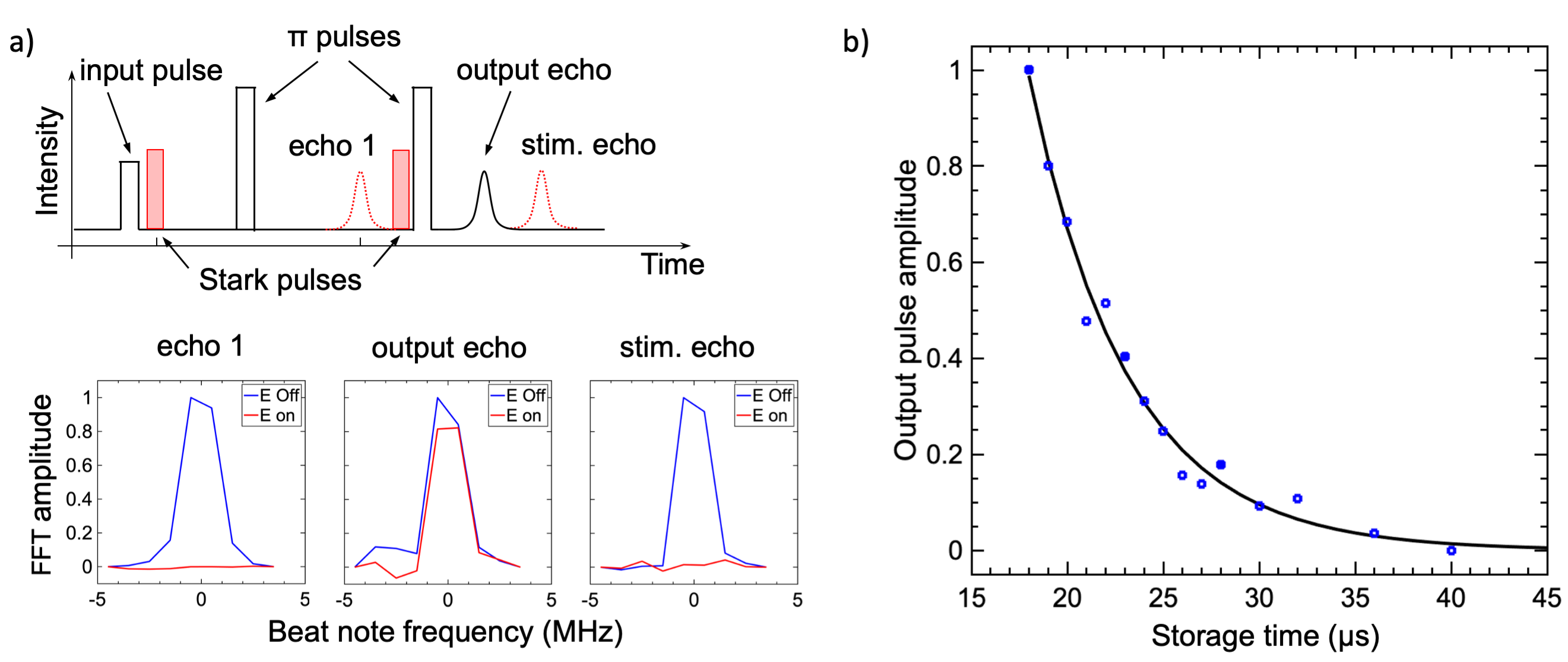}
  \caption{Coherent storage in nanoparticles. (a) Upper panel: Pulse sequence for the Stark Echo Envelope Modulation Memory (SEMM) protocol (black: light, red: electric field). Lower panel: Fourier transform of the heterodyne detected echoes showing the strong cancellation of echo 1 and  stimulated echo by the Stark pulses, as well as the output echo revival (see text). The frequency origin is set at the heterodyne beat note (30 MHz). The FFT resolution is limited by the $\approx$ 1 $\mu$s length of the echoes. (b) Output pulse amplitude in the SEMM sequence as a function of the total storage time. Circles: experimental data, line: single exponential fit. Effective storage time is $T_{2,M} = 5.1 \pm$ 0.5 $\mu$s.}
  \label{fig: fig3}
\end{figure}

Fig. \ref{fig: fig2} shows that \eu\ ions optical phase can be precisely controlled over a large bandwidth by an electric field in nanoparticles. This is the basis for the storage experiments we present now. The pulse sequence of the SEMM protocol is sketched in Fig. \ref{fig: fig3}a. Optical rephasing of the input pulse is achieved by two $\pi$ pulses in order that the final echo, i.e. the output pulse, is emitted in a non-inverted medium. Fluorescence noise is suppressed, allowing the protocol to be used with photonic qubits such as single photons \cite{Ruggiero:2009uu}. An important part of the protocol is to suppress the intermediate echo (echo 1) that appears after the first $\pi$ pulse. A Stark pulse producing a $\pi$/2 phase shift can achieve this. The same pulse is applied a second time to reverse the initial phase shift and recover the output echo with a high intensity. It also suppresses the echo due to spontaneous photons emitted at the same time as echo 1 and that would therefore be rephased at the same time as the output echo \cite{Arcangeli:2016eu}. The delays between input and rephasing pulses are chosen to allow separating the output pulse from the stimulated echo that can be emitted because of imperfect $\pi$ pulses.

Experimentally, optical pulse  amplitudes and lengths were tuned for maximal output echo signal (see SI). Given the strongly scattering medium, Rabi frequencies greatly vary across the sample and pulse area are not well defined. This prevents determining $\pi$ pulses from Rabi oscillations for example. We then set the Stark pulse area to a value of 10 V/cm$\cdot\mu$s in order to cancel  echo 1 emission.  The Fourier transform of the heterodyne detection pulse shows an attenuation of echo 1 by a factor $> 100$ in amplitude, or 10$^4$ in intensity (Fig. \ref{fig: fig3}a). Such an attenuation would be compatible with operation in the few photon regime, assuming high fidelity optical $\pi$ pulses \cite{Arcangeli:2016eu}. In opposition to echo 1, the output pulse is well recovered after the second Stark pulse. No significant difference in amplitude is observed with or without electric field because the echo emissions have a low efficiency and only a small fraction of ions are involved in echo 1 emission. We also note that the stimulated echo is strongly attenuated by the Stark pulses. This is expected since the second Stark pulse has no effect on the populations from which the stimulated echo arises and cannot recover the phase shifts created by the first Stark pulse. 

Storage time was determined by varying the delay between the input  and the first $\pi$ pulses, leading to an effective coherence lifetime of $T_{2,M} = 5.1 \pm 0.5$ $\mu$s for the memory, in good agreement with the photon echo $T_2 = 5.7 \pm 0.2$ $\mu$s. This value is more than two orders of magnitude longer than the one  obtained in a nano-structured bulk crystal, although these previous results were obtained at the single photon level \cite{Zhong:2017fe}. We also note that SEMM allows for on-demand retrieval  without spin storage. Output signals could be observed until about 40 $\mu$s storage time, corresponding to a retrieval time adjustable within 22 $\mu$s.

\begin{figure}
  \includegraphics[width =0.7\textwidth]{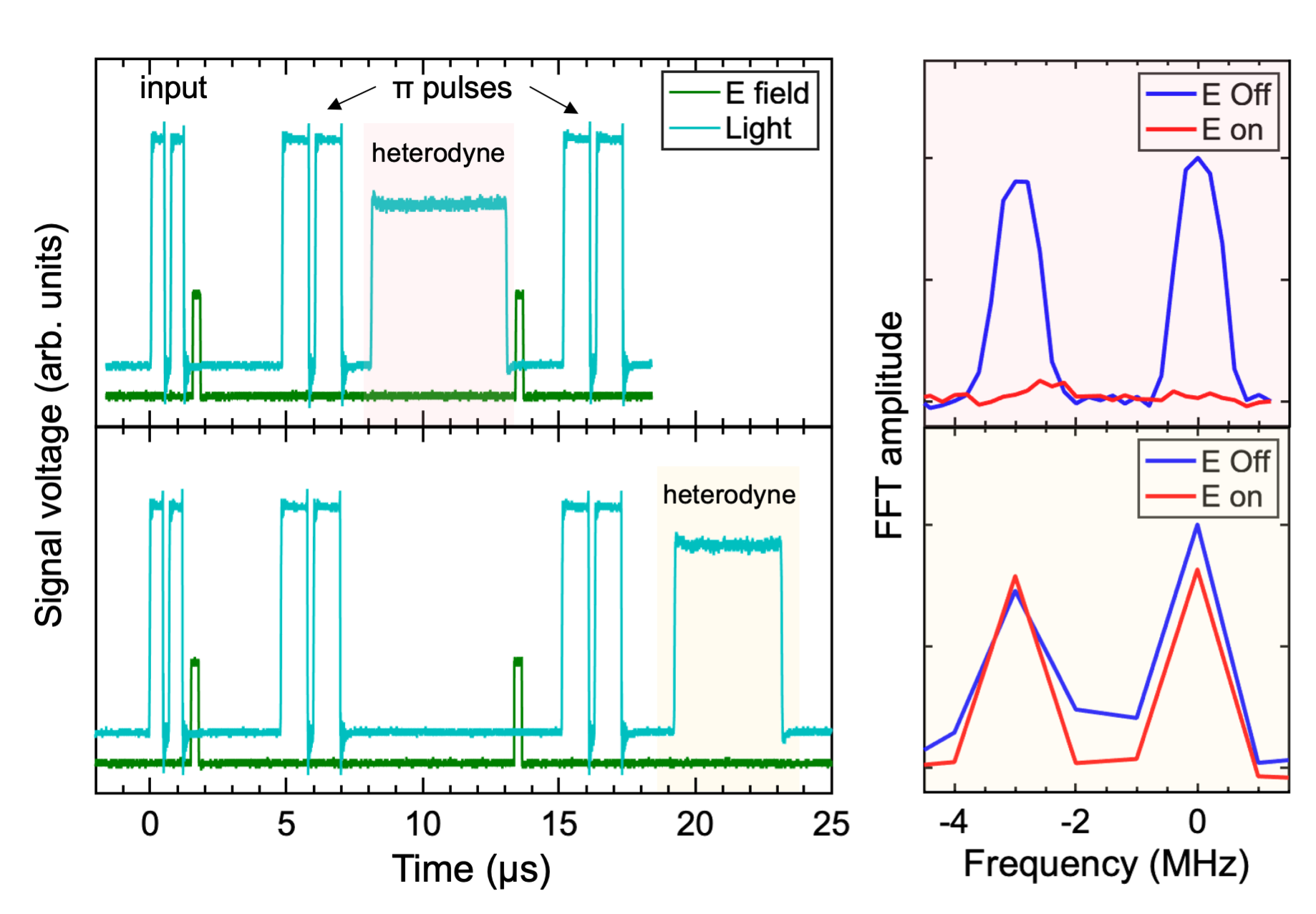}
  \caption{Frequency-multiplexed storage with SEMM protocol. Left panel: photo-detector and voltage source signals during storage of two pulses with frequencies separated by 3 MHz. The broad heterodyne pulse detects the two echoes 1 (upper graph) and the two output echoes (lower graph). Right panel: Normalized Fourier transform of the heterodyne detected echoes with or without Stark pulses. The Stark pulses cancel the two echoes 1 and restore the output echoes. Frequency origin is set at the heterodyne beat note with the highest frequency echo (30 MHz).}
  \label{fig: fig4}
\end{figure}

The small variation of the Stark coefficient with optical frequency suggests that SEMM storage could be obtained over broad bandwidths. First, we verified that strong cancellation of intermediate echoes and efficient revival of output pulses was simultaneously possible at two frequencies. As seen in Fig. \ref{fig: fig4}, two successive input pulses, with frequencies separated by 3 MHz, were stored  and then retrieved by two pairs of $\pi$ pulses also separated by 3 MHz. Echoes were detected by a single heterodyne pulse long enough to probe both echoes. When Stark pulses with the same area than in the single frequency storage are turned on, a strong attenuation is achieved for echoes 1, whereas the output echoes are well retrieved, as seen in the FFT of the heterodyne pulses (Fig. \ref{fig: fig4}). This confirms the possibility of frequency-multiplexed storage, although the experiment should be reproduced with larger frequency differences or shorter pulses. 

\begin{figure}
  \includegraphics[width = \textwidth]{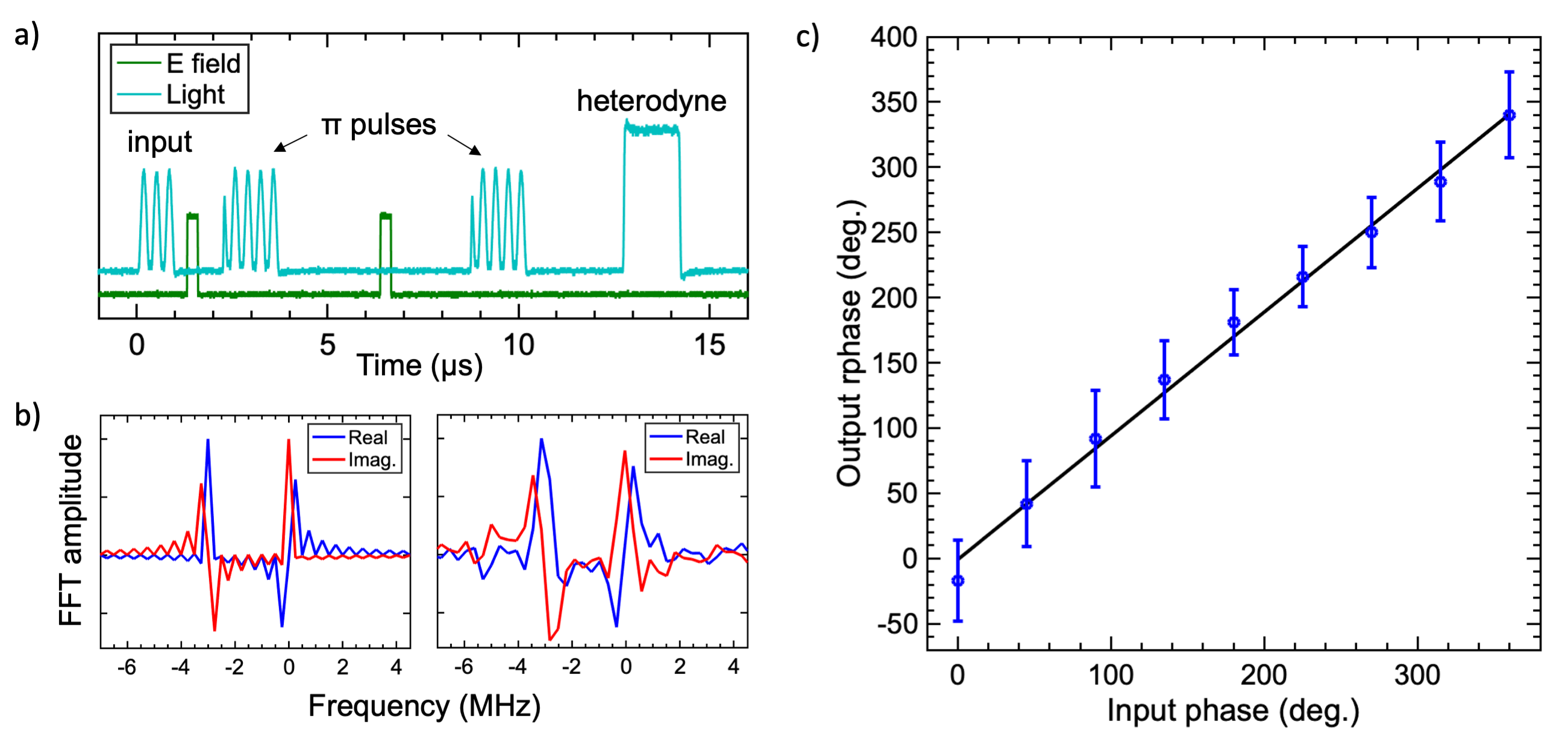}
  \caption{SEMM storage fidelity. (a) Photo-detector and voltage source signals for storage of a two-color pulse (3 MHz frequency difference). The heterodyne pulse detects the output echo. (b) Simulated (left) and experimental (right) Fourier transform of the heterodyne detected output echo for a two-color input pulse with a $\pi/2$ relative phase shift. (c) Output vs. input phase shifts in two-color storage. Circles: experimental data, line: linear fit, correlation coefficient: 0.997.}
  \label{fig: fig5}
\end{figure}

We finally used multi-frequency storage to probe memory fidelity for phase qubits. Since our laser has a coherence time of only about 1 $\mu$s, it cannot directly provide a phase reference. To circumvent this limitation, two-color laser pulses instead of successive pulses of different frequencies were stored. Generating the pulses in a single acousto-optic modulator and coupling the overlapping diffracted beams in a single-mode fiber gave an adjustable and highly stable relative phase between the two frequencies. The output echo is again detected by a heterodyne pulse and the relative phases between the two-colors, separated by 3 MHz, analyzed by Fourier transform. The sequence is shown in Fig. \ref{fig: fig5}. The two-color pulses show a strong intensity modulation confirming the relative phase stability since these traces are averaged over $\approx$ 1 s. Phase difference was set to zero for the two-color $\pi$ pulses, and only input pulse relative phase was varied.

The echo signals were clearly observed, but with  lower signal to noise ratios (SNR) than for storage of successive pulses. This is due to the low efficiency of the simultaneous two-color generation. Fig. \ref{fig: fig5}b shows simulation and experimental data for the real and imaginary parts of the heterodyne pulse FFT in the case of a $\pi/2$ relative phase shift. As the heterodyne pulse adds a random phase, it was necessary to increase SNR by averaging about 40 single-shot experimental traces showing similar patterns, i.e. a positive real part at -3 MHz. A complete analysis was performed by calculating phases at 0 and -3 MHz  from the amplitudes of FFT real and imaginary parts  and averaging their differences over all available data. The latter amounted to 200 traces for each relative phase (see SI). A high correlation between input and output phases was found, as displayed in Fig. \ref{fig: fig5}c. Although this result is not unexpected in echo based protocols, it nevertheless demonstrates the high phase fidelity of the memory, a fundamental requirement for quantum storage.

We now briefly discuss how the results presented in this Letter can be applied to a single particle inserted in an optical cavity. A high finesse, low volume cavity would first allow for high storage and retrieval efficiency by increasing light-ion coupling \cite{Casabone:2018kc}. The SEMM protocol would be particularly well adapted to this case as it does not rely on spatial phase matching \cite{Damon:2011tx} for intermediate echo suppression. In case of very high light-ion coupling, it may  also be necessary to reduce the cavity finesse during the storage time. This will reduce the spontaneous emission that could be rephased at the same time as the memory output and superradiant emissions \cite{Afzelius:2013ga,Julsgaard:2013br,Arcangeli:2016eu}. The fast tunability of the fiber cavity would be very useful in this case \cite{Casabone:2020wj}. 
Only very small voltages would be needed to implement SEMM frequency shifts since a single particle can sit very close to electrodes. The cavity can ensure high Rabi frequencies across a broad frequency range and with a high spatial homogeneity. This will enable $\pi$ pulses with much higher fidelity than in experiments using bulk crystals, a key requirement for low-noise operation in the quantum regime. Longer storage times could be obtained by inserting high fidelity transfer pulses to nuclear spin states in the SEMM sequence. Since  spin coherence lifetimes can reach several ms in these nanoparticles \cite{Serrano:2018ea}, storage time would be extended  by three orders of magnitude. The relatively small number of ions in a single particle, $3 \times 10^4$  in the present 100 nm diameter particles ($C_2$ site), also opens the prospect of single spin processing for entanglement purification or swapping \cite{Kindem:2020eb,Raha:2020jb}. Spins would be controlled using 2-color selective optical excitations \cite{Walther:2015hh,Serrano:2018ea}, taking advantage of the large inhomogeneous over homogeneous linewidth ratio in the nanoparticles ($\approx 5 \times 10^5$). Applying electric fields to Stark shift optical transition would then provide an additional control parameter to implement qubit gates \cite{Wolfowicz:2014ji}. 

In conclusion, a quantum memory protocol for light based on the linear Stark effect has been demonstrated in \eu:\YO\ doped nanoparticles. We measured a Stark coefficient for the $^5$D$_0$ -$^7$F$_0$ transition of $49.5 \pm 1$ kHz/(V/cm), which shows variations lower than a few kHz over the full width of the  optical inhomogeneous linewidth. Using suitable electric field pulses, on-demand optical storage was achieved, with strong suppression of noisy signals while maintaining output compatible with quantum operation. The memory effective coherence lifetime is $5.1 \pm 0.5$ $\mu$s.
Frequency-multiplexed storage was demonstrated with similar properties, confirming the small variations of Stark coefficients and enabling observation of a high phase fidelity between input and output pulses. 
These results open the way to using RE doped nanoparticles  as efficient light-interfaces for quantum networks and other solid-state optical quantum technologies.

\begin{acknowledgement}
Alexandre Fossati and Shuping Liu equally contributed to this work.
This project has received funding from the European Union’s Horizon 2020 research and innovation programme under grant agreements No 712721 (NanOQTech) and 820391 (SQUARE). S. Liu is supported by the Key R\&D Program of Guangdong province (Grant No. 2018B030325001).

\end{acknowledgement}

\newpage
\bibliography{refs}

\providecommand{\latin}[1]{#1}
\makeatletter
\providecommand{\doi}
  {\begingroup\let\do\@makeother\dospecials
  \catcode`\{=1 \catcode`\}=2 \doi@aux}
\providecommand{\doi@aux}[1]{\endgroup\texttt{#1}}
\makeatother
\providecommand*\mcitethebibliography{\thebibliography}
\csname @ifundefined\endcsname{endmcitethebibliography}
  {\let\endmcitethebibliography\endthebibliography}{}
\begin{mcitethebibliography}{44}
\providecommand*\natexlab[1]{#1}
\providecommand*\mciteSetBstSublistMode[1]{}
\providecommand*\mciteSetBstMaxWidthForm[2]{}
\providecommand*\mciteBstWouldAddEndPuncttrue
  {\def\EndOfBibitem{\unskip.}}
\providecommand*\mciteBstWouldAddEndPunctfalse
  {\let\EndOfBibitem\relax}
\providecommand*\mciteSetBstMidEndSepPunct[3]{}
\providecommand*\mciteSetBstSublistLabelBeginEnd[3]{}
\providecommand*\EndOfBibitem{}
\mciteSetBstSublistMode{f}
\mciteSetBstMaxWidthForm{subitem}{(\alph{mcitesubitemcount})}
\mciteSetBstSublistLabelBeginEnd
  {\mcitemaxwidthsubitemform\space}
  {\relax}
  {\relax}

\bibitem[Goldner \latin{et~al.}(2015)Goldner, Ferrier, and
  Guillot-No{\"e}l]{Goldner:2015ve}
Goldner,~P.; Ferrier,~A.; Guillot-No{\"e}l,~O. In \emph{Handbook on the Physics
  and Chemistry of Rare Earths}; B{\"u}nzli,~J.-C.~G., Pecharsky,~V.~K., Eds.;
  Elsevier: Amsterdam, 2015; pp 1--78\relax
\mciteBstWouldAddEndPuncttrue
\mciteSetBstMidEndSepPunct{\mcitedefaultmidpunct}
{\mcitedefaultendpunct}{\mcitedefaultseppunct}\relax
\EndOfBibitem
\bibitem[Thiel \latin{et~al.}(2011)Thiel, B{\"o}ttger, and Cone]{Thiel:2011eu}
Thiel,~C.~W.; B{\"o}ttger,~T.; Cone,~R.~L. {Rare-earth-doped materials for
  applications in quantum information storage and signal processing}. \emph{J.
  Lumin.} \textbf{2011}, \emph{131}, 353--361\relax
\mciteBstWouldAddEndPuncttrue
\mciteSetBstMidEndSepPunct{\mcitedefaultmidpunct}
{\mcitedefaultendpunct}{\mcitedefaultseppunct}\relax
\EndOfBibitem
\bibitem[Equall \latin{et~al.}(1994)Equall, Sun, Cone, and
  Macfarlane]{Equall:1994dn}
Equall,~R.~W.; Sun,~Y.; Cone,~R.~L.; Macfarlane,~R.~M. {Ultraslow optical
  dephasing in Eu$^{3+}$:Y$_{2}$SiO$_{5}$}. \emph{Phys. Rev. Lett.}
  \textbf{1994}, \emph{72}, 2179--2182\relax
\mciteBstWouldAddEndPuncttrue
\mciteSetBstMidEndSepPunct{\mcitedefaultmidpunct}
{\mcitedefaultendpunct}{\mcitedefaultseppunct}\relax
\EndOfBibitem
\bibitem[B{\"o}ttger \latin{et~al.}(2009)B{\"o}ttger, Thiel, Cone, and
  Sun]{Bottger:2009ik}
B{\"o}ttger,~T.; Thiel,~C.~W.; Cone,~R.~L.; Sun,~Y. {Effects of magnetic field
  orientation on optical decoherence in Er$^{3+}$:Y$_{2}$SiO$_{5}$}.
  \emph{Phys. Rev. B} \textbf{2009}, \emph{79}, 115104\relax
\mciteBstWouldAddEndPuncttrue
\mciteSetBstMidEndSepPunct{\mcitedefaultmidpunct}
{\mcitedefaultendpunct}{\mcitedefaultseppunct}\relax
\EndOfBibitem
\bibitem[Businger \latin{et~al.}(2020)Businger, Tiranov, Kaczmarek, Welinski,
  Zhang, Ferrier, Goldner, and Afzelius]{Businger:2020jf}
Businger,~M.; Tiranov,~A.; Kaczmarek,~K.~T.; Welinski,~S.; Zhang,~Z.;
  Ferrier,~A.; Goldner,~P.; Afzelius,~M. {Optical Spin-Wave Storage in a
  Solid-State Hybridized Electron-Nuclear Spin Ensemble}. \emph{Phys. Rev.
  Lett.} \textbf{2020}, \emph{124}, 053606\relax
\mciteBstWouldAddEndPuncttrue
\mciteSetBstMidEndSepPunct{\mcitedefaultmidpunct}
{\mcitedefaultendpunct}{\mcitedefaultseppunct}\relax
\EndOfBibitem
\bibitem[Zhong \latin{et~al.}(2015)Zhong, Hedges, Ahlefeldt, Bartholomew,
  Beavan, Wittig, Longdell, and Sellars]{Zhong:2015bw}
Zhong,~M.; Hedges,~M.~P.; Ahlefeldt,~R.~L.; Bartholomew,~J.~G.; Beavan,~S.~E.;
  Wittig,~S.~M.; Longdell,~J.~J.; Sellars,~M.~J. {Optically addressable nuclear
  spins in a solid with a six-hour coherence time}. \emph{Nature}
  \textbf{2015}, \emph{517}, 177--180\relax
\mciteBstWouldAddEndPuncttrue
\mciteSetBstMidEndSepPunct{\mcitedefaultmidpunct}
{\mcitedefaultendpunct}{\mcitedefaultseppunct}\relax
\EndOfBibitem
\bibitem[Lvovsky \latin{et~al.}(2009)Lvovsky, Sanders, and
  Tittel]{Lvovsky:2009fr}
Lvovsky,~A.~I.; Sanders,~B.~C.; Tittel,~W. {Optical quantum memory}. \emph{Nat.
  Photonics} \textbf{2009}, \emph{3}, 706--714\relax
\mciteBstWouldAddEndPuncttrue
\mciteSetBstMidEndSepPunct{\mcitedefaultmidpunct}
{\mcitedefaultendpunct}{\mcitedefaultseppunct}\relax
\EndOfBibitem
\bibitem[Afzelius and de~Riedmatten(2015)Afzelius, and
  de~Riedmatten]{deRiedmatten:2015tj}
Afzelius,~M.; de~Riedmatten,~H. In \emph{Engineering the Atom-Photon
  Interaction}; Predojevi{\'c},~A., Mitchell,~M.~W., Eds.; Springer, 2015\relax
\mciteBstWouldAddEndPuncttrue
\mciteSetBstMidEndSepPunct{\mcitedefaultmidpunct}
{\mcitedefaultendpunct}{\mcitedefaultseppunct}\relax
\EndOfBibitem
\bibitem[Awschalom \latin{et~al.}(2018)Awschalom, Hanson, Wrachtrup, and
  Zhou]{Awschalom:2018ic}
Awschalom,~D.~D.; Hanson,~R.; Wrachtrup,~J.; Zhou,~B.~B. {Quantum technologies
  with optically interfaced solid-state spins}. \emph{Nat. Photonics}
  \textbf{2018}, \emph{12}, 1--12\relax
\mciteBstWouldAddEndPuncttrue
\mciteSetBstMidEndSepPunct{\mcitedefaultmidpunct}
{\mcitedefaultendpunct}{\mcitedefaultseppunct}\relax
\EndOfBibitem
\bibitem[Hedges \latin{et~al.}(2010)Hedges, Longdell, Li, and
  Sellars]{Hedges:2010dq}
Hedges,~M.~P.; Longdell,~J.~J.; Li,~Y.; Sellars,~M.~J. {Efficient quantum
  memory for light}. \emph{Nature} \textbf{2010}, \emph{465}, 1052--1056\relax
\mciteBstWouldAddEndPuncttrue
\mciteSetBstMidEndSepPunct{\mcitedefaultmidpunct}
{\mcitedefaultendpunct}{\mcitedefaultseppunct}\relax
\EndOfBibitem
\bibitem[Holz{\"a}pfel \latin{et~al.}(2019)Holz{\"a}pfel, Etesse, Kaczmarek,
  Tiranov, Gisin, and Afzelius]{Holzapfel:2019ks}
Holz{\"a}pfel,~A.; Etesse,~J.; Kaczmarek,~K.~T.; Tiranov,~A.; Gisin,~N.;
  Afzelius,~M. {Optical storage for 0.53 seconds in a solid-state atomic
  frequency comb memory using dynamical decoupling}. \emph{New J. Phys.}
  \textbf{2019}, \relax
\mciteBstWouldAddEndPunctfalse
\mciteSetBstMidEndSepPunct{\mcitedefaultmidpunct}
{}{\mcitedefaultseppunct}\relax
\EndOfBibitem
\bibitem[Kindem \latin{et~al.}(2020)Kindem, Ruskuc, Bartholomew, Rochman, Huan,
  and Faraon]{Kindem:2020eb}
Kindem,~J.~M.; Ruskuc,~A.; Bartholomew,~J.~G.; Rochman,~J.; Huan,~Y.~Q.;
  Faraon,~A. {Control and single-shot readout of an ion embedded in a
  nanophotonic cavity}. \emph{Nature} \textbf{2020}, \emph{580}, 1--12\relax
\mciteBstWouldAddEndPuncttrue
\mciteSetBstMidEndSepPunct{\mcitedefaultmidpunct}
{\mcitedefaultendpunct}{\mcitedefaultseppunct}\relax
\EndOfBibitem
\bibitem[Raha \latin{et~al.}(2020)Raha, Chen, Phenicie, Ourari, Dibos, and
  Thompson]{Raha:2020jb}
Raha,~M.; Chen,~S.; Phenicie,~C.~M.; Ourari,~S.; Dibos,~A.~M.; Thompson,~J.~D.
  {Optical quantum nondemolition measurement of a single rare earth ion qubit}.
  \emph{Nat. Commun.} \textbf{2020}, \emph{11}, 1--6\relax
\mciteBstWouldAddEndPuncttrue
\mciteSetBstMidEndSepPunct{\mcitedefaultmidpunct}
{\mcitedefaultendpunct}{\mcitedefaultseppunct}\relax
\EndOfBibitem
\bibitem[Zhong and Goldner(2019)Zhong, and Goldner]{Zhong:2019gf}
Zhong,~T.; Goldner,~P. {Emerging rare-earth doped material platforms for
  quantum nanophotonics}. \emph{Nanophotonics} \textbf{2019}, \emph{8},
  2003--2015\relax
\mciteBstWouldAddEndPuncttrue
\mciteSetBstMidEndSepPunct{\mcitedefaultmidpunct}
{\mcitedefaultendpunct}{\mcitedefaultseppunct}\relax
\EndOfBibitem
\bibitem[Scarafagio \latin{et~al.}(2020)Scarafagio, Tallaire, Chavanne, Cassir,
  Ringued{\'e}, Serrano, Goldner, and Ferrier]{Scarafagio:2020kl}
Scarafagio,~M.; Tallaire,~A.; Chavanne,~M.-H.; Cassir,~M.; Ringued{\'e},~A.;
  Serrano,~D.; Goldner,~P.; Ferrier,~A. {Improving the Luminescent Properties
  of Atomic Layer Deposition Eu:Y 2O 3Thin Films through Optimized Thermal
  Annealing}. \emph{Phys. Stat. Sol. (a)} \textbf{2020}, \emph{46},
  1900909--7\relax
\mciteBstWouldAddEndPuncttrue
\mciteSetBstMidEndSepPunct{\mcitedefaultmidpunct}
{\mcitedefaultendpunct}{\mcitedefaultseppunct}\relax
\EndOfBibitem
\bibitem[Singh \latin{et~al.}(2020)Singh, Prakash, Wolfowicz, Wen, Huang, Rajh,
  Awschalom, Zhong, and Guha]{Singh:2020ey}
Singh,~M.~K.; Prakash,~A.; Wolfowicz,~G.; Wen,~J.; Huang,~Y.; Rajh,~T.;
  Awschalom,~D.~D.; Zhong,~T.; Guha,~S. {Epitaxial Er-doped Y2O3 on silicon for
  quantum coherent devices}. \emph{APL Mater.} \textbf{2020}, \emph{8},
  031111\relax
\mciteBstWouldAddEndPuncttrue
\mciteSetBstMidEndSepPunct{\mcitedefaultmidpunct}
{\mcitedefaultendpunct}{\mcitedefaultseppunct}\relax
\EndOfBibitem
\bibitem[Dutta \latin{et~al.}(2019)Dutta, Goldschmidt, Barik, Saha, and
  Waks]{Dutta:2019fk}
Dutta,~S.; Goldschmidt,~E.~A.; Barik,~S.; Saha,~U.; Waks,~E. {Integrated
  Photonic Platform for Rare-Earth Ions in Thin Film Lithium Niobate}.
  \emph{Nano. Lett.} \textbf{2019}, \emph{20}, 741--747\relax
\mciteBstWouldAddEndPuncttrue
\mciteSetBstMidEndSepPunct{\mcitedefaultmidpunct}
{\mcitedefaultendpunct}{\mcitedefaultseppunct}\relax
\EndOfBibitem
\bibitem[Bartholomew \latin{et~al.}(2017)Bartholomew, de~Oliveira~Lima,
  Ferrier, and Goldner]{Bartholomew:2017ik}
Bartholomew,~J.~G.; de~Oliveira~Lima,~K.; Ferrier,~A.; Goldner,~P. {Optical
  Line Width Broadening Mechanisms at the 10 kHz Level in Eu3+:Y2O3
  Nanoparticles}. \emph{Nano. Lett.} \textbf{2017}, \emph{17}, 778--787\relax
\mciteBstWouldAddEndPuncttrue
\mciteSetBstMidEndSepPunct{\mcitedefaultmidpunct}
{\mcitedefaultendpunct}{\mcitedefaultseppunct}\relax
\EndOfBibitem
\bibitem[Serrano \latin{et~al.}(2018)Serrano, Karlsson, Fossati, Ferrier, and
  Goldner]{Serrano:2018ea}
Serrano,~D.; Karlsson,~J.; Fossati,~A.; Ferrier,~A.; Goldner,~P. {All-optical
  control of long-lived nuclear spins in rare-earth doped nanoparticles}.
  \emph{Nat. Commun.} \textbf{2018}, \emph{9}, 2127\relax
\mciteBstWouldAddEndPuncttrue
\mciteSetBstMidEndSepPunct{\mcitedefaultmidpunct}
{\mcitedefaultendpunct}{\mcitedefaultseppunct}\relax
\EndOfBibitem
\bibitem[Serrano \latin{et~al.}(2019)Serrano, Deshmukh, Liu, Tallaire, Ferrier,
  de~Riedmatten, and Goldner]{Serrano:2019km}
Serrano,~D.; Deshmukh,~C.; Liu,~S.; Tallaire,~A.; Ferrier,~A.;
  de~Riedmatten,~H.; Goldner,~P. {Coherent optical and spin spectroscopy of
  nanoscale Pr$^{3+}$:Y$_{2}$O$_{3}$}. \emph{Phys. Rev. B} \textbf{2019},
  \emph{100}, 144304\relax
\mciteBstWouldAddEndPuncttrue
\mciteSetBstMidEndSepPunct{\mcitedefaultmidpunct}
{\mcitedefaultendpunct}{\mcitedefaultseppunct}\relax
\EndOfBibitem
\bibitem[Liu \latin{et~al.}(2018)Liu, Serrano, Fossati, Tallaire, Ferrier, and
  Goldner]{Liu:2018gk}
Liu,~S.; Serrano,~D.; Fossati,~A.; Tallaire,~A.; Ferrier,~A.; Goldner,~P.
  {Controlled size reduction of rare earth doped nanoparticles for optical
  quantum technologies}. \emph{RSC Adv.} \textbf{2018}, \emph{8},
  37098--37104\relax
\mciteBstWouldAddEndPuncttrue
\mciteSetBstMidEndSepPunct{\mcitedefaultmidpunct}
{\mcitedefaultendpunct}{\mcitedefaultseppunct}\relax
\EndOfBibitem
\bibitem[Casabone \latin{et~al.}(2018)Casabone, Benedikter, H{\"u}mmer, Oehl,
  de~Oliveira~Lima, H{\"a}nsch, Ferrier, Goldner, de~Riedmatten, and
  Hunger]{Casabone:2018kc}
Casabone,~B.; Benedikter,~J.; H{\"u}mmer,~T.; Oehl,~F.; de~Oliveira~Lima,~K.;
  H{\"a}nsch,~T.~W.; Ferrier,~A.; Goldner,~P.; de~Riedmatten,~H.; Hunger,~D.
  {Cavity-enhanced spectroscopy of a few-ion ensemble in Eu3+:Y2O3}. \emph{New
  J. Phys.} \textbf{2018}, \emph{20}, 095006--9\relax
\mciteBstWouldAddEndPuncttrue
\mciteSetBstMidEndSepPunct{\mcitedefaultmidpunct}
{\mcitedefaultendpunct}{\mcitedefaultseppunct}\relax
\EndOfBibitem
\bibitem[Casabone \latin{et~al.}(2020)Casabone, Deshmukh, Liu, Serrano,
  Ferrier, H{\"u}mmer, Goldner, Hunger, and de~Riedmatten]{Casabone:2020wj}
Casabone,~B.; Deshmukh,~C.; Liu,~S.; Serrano,~D.; Ferrier,~A.; H{\"u}mmer,~T.;
  Goldner,~P.; Hunger,~D.; de~Riedmatten,~H. {Dynamic control of Purcell
  enhanced emission of erbium ions in nanoparticles}. \emph{arXiv}
  \textbf{2020}, \relax
\mciteBstWouldAddEndPunctfalse
\mciteSetBstMidEndSepPunct{\mcitedefaultmidpunct}
{}{\mcitedefaultseppunct}\relax
\EndOfBibitem
\bibitem[Arcangeli \latin{et~al.}(2016)Arcangeli, Ferrier, and
  Goldner]{Arcangeli:2016eu}
Arcangeli,~A.; Ferrier,~A.; Goldner,~P. {Stark echo modulation for quantum
  memories}. \emph{Phys. Rev. A} \textbf{2016}, \emph{93}, 062303\relax
\mciteBstWouldAddEndPuncttrue
\mciteSetBstMidEndSepPunct{\mcitedefaultmidpunct}
{\mcitedefaultendpunct}{\mcitedefaultseppunct}\relax
\EndOfBibitem
\bibitem[Damon \latin{et~al.}(2011)Damon, Bonarota, Louchet-Chauvet,
  Chaneli{\`e}re, and Le~Gou{\"e}t]{Damon:2011tx}
Damon,~V.; Bonarota,~M.; Louchet-Chauvet,~A.; Chaneli{\`e}re,~T.;
  Le~Gou{\"e}t,~J.-L. {Revival of silenced echo and quantum memory for light}.
  \emph{New J. Phys.} \textbf{2011}, \emph{13}, 093031\relax
\mciteBstWouldAddEndPuncttrue
\mciteSetBstMidEndSepPunct{\mcitedefaultmidpunct}
{\mcitedefaultendpunct}{\mcitedefaultseppunct}\relax
\EndOfBibitem
\bibitem[McAuslan \latin{et~al.}(2011)McAuslan, Ledingham, Naylor, Beavan,
  Hedges, Sellars, and Longdell]{McAuslan:2011ke}
McAuslan,~D.~L.; Ledingham,~P.~M.; Naylor,~W.~R.; Beavan,~S.~E.; Hedges,~M.~P.;
  Sellars,~M.~J.; Longdell,~J.~J. {Photon-echo quantum memories in
  inhomogeneously broadened two-level atoms}. \emph{Phys. Rev. A}
  \textbf{2011}, \emph{84}, 022309\relax
\mciteBstWouldAddEndPuncttrue
\mciteSetBstMidEndSepPunct{\mcitedefaultmidpunct}
{\mcitedefaultendpunct}{\mcitedefaultseppunct}\relax
\EndOfBibitem
\bibitem[Afzelius \latin{et~al.}(2013)Afzelius, Sangouard, Johansson, Staudt,
  and Wilson]{Afzelius:2013ga}
Afzelius,~M.; Sangouard,~N.; Johansson,~G.; Staudt,~M.~U.; Wilson,~C.~M.
  {Proposal for a coherent quantum memory for propagating microwave photons}.
  \emph{New J. Phys.} \textbf{2013}, \emph{15}, 065008\relax
\mciteBstWouldAddEndPuncttrue
\mciteSetBstMidEndSepPunct{\mcitedefaultmidpunct}
{\mcitedefaultendpunct}{\mcitedefaultseppunct}\relax
\EndOfBibitem
\bibitem[Macfarlane(2007)]{Macfarlane:2007it}
Macfarlane,~R.~M. {Optical Stark spectroscopy of solids}. \emph{J. Lumin.}
  \textbf{2007}, \emph{125}, 156--174\relax
\mciteBstWouldAddEndPuncttrue
\mciteSetBstMidEndSepPunct{\mcitedefaultmidpunct}
{\mcitedefaultendpunct}{\mcitedefaultseppunct}\relax
\EndOfBibitem
\bibitem[Kaplyanskii(2002)]{Kaplyanskii:2002cy}
Kaplyanskii,~A.~A. {Linear Stark effect in spectroscopy and luminescence of
  doped inorganic insulating crystals}. \emph{J. Lumin.} \textbf{2002},
  \emph{100}, 21--34\relax
\mciteBstWouldAddEndPuncttrue
\mciteSetBstMidEndSepPunct{\mcitedefaultmidpunct}
{\mcitedefaultendpunct}{\mcitedefaultseppunct}\relax
\EndOfBibitem
\bibitem[Graf \latin{et~al.}(1997)Graf, Renn, Wild, and Mitsunaga]{Graf:1997kn}
Graf,~F.~R.; Renn,~A.; Wild,~U.~P.; Mitsunaga,~M. {Site interference in
  Stark-modulated photon echoes }. \emph{Phys. Rev. B} \textbf{1997},
  \emph{55}, 11225--11229\relax
\mciteBstWouldAddEndPuncttrue
\mciteSetBstMidEndSepPunct{\mcitedefaultmidpunct}
{\mcitedefaultendpunct}{\mcitedefaultseppunct}\relax
\EndOfBibitem
\bibitem[Liu \latin{et~al.}(2020)Liu, Fossati, Serrano, Tallaire, Ferrier, and
  Goldner]{Liu:2020ty}
Liu,~S.; Fossati,~A.; Serrano,~D.; Tallaire,~A.; Ferrier,~A.; Goldner,~P.
  {Defect Engineering for Quantum Grade Rare-Earth Nanocrystals}.
  \emph{ChemRxiv} \textbf{2020}, 1--33\relax
\mciteBstWouldAddEndPuncttrue
\mciteSetBstMidEndSepPunct{\mcitedefaultmidpunct}
{\mcitedefaultendpunct}{\mcitedefaultseppunct}\relax
\EndOfBibitem
\bibitem[Perrot \latin{et~al.}(2013)Perrot, Goldner, Giaume, Lovri{\'c},
  Andriamiadamanana, Gon{\c c}alves, and Ferrier]{Perrot:2013hy}
Perrot,~A.; Goldner,~P.; Giaume,~D.; Lovri{\'c},~M.; Andriamiadamanana,~C.;
  Gon{\c c}alves,~R.~R.; Ferrier,~A. {Narrow Optical Homogeneous Linewidths in
  Rare Earth Doped Nanocrystals}. \emph{Phys. Rev. Lett.} \textbf{2013},
  \emph{111}, 203601\relax
\mciteBstWouldAddEndPuncttrue
\mciteSetBstMidEndSepPunct{\mcitedefaultmidpunct}
{\mcitedefaultendpunct}{\mcitedefaultseppunct}\relax
\EndOfBibitem
\bibitem[Mims(1964)]{Mims:1964cj}
Mims,~W.~B. {Electric Field Effects in Spin Echoes}. \emph{Phys. Rev.}
  \textbf{1964}, \emph{133}, A835--A840\relax
\mciteBstWouldAddEndPuncttrue
\mciteSetBstMidEndSepPunct{\mcitedefaultmidpunct}
{\mcitedefaultendpunct}{\mcitedefaultseppunct}\relax
\EndOfBibitem
\bibitem[Meixner \latin{et~al.}(1992)Meixner, Jefferson, and
  Macfarlane]{Meixner:1992uu}
Meixner,~A.~J.; Jefferson,~C.~M.; Macfarlane,~R.~M. {Measurement of the Stark
  effect with subhomogeneous linewidth resolution in Eu 3+: YAlO 3 with the use
  of photon-echo modulation}. \emph{Phys. Rev. B} \textbf{1992}, \emph{46},
  5912--5916\relax
\mciteBstWouldAddEndPuncttrue
\mciteSetBstMidEndSepPunct{\mcitedefaultmidpunct}
{\mcitedefaultendpunct}{\mcitedefaultseppunct}\relax
\EndOfBibitem
\bibitem[Macfarlane \latin{et~al.}(2014)Macfarlane, Arcangeli, Ferrier, and
  Goldner]{Macfarlane:2014fy}
Macfarlane,~R.~M.; Arcangeli,~A.; Ferrier,~A.; Goldner,~P. {Optical Measurement
  of the Effect of Electric Fields on the Nuclear Spin Coherence of Rare-Earth
  Ions in Solids}. \emph{Phys. Rev. Lett.} \textbf{2014}, \emph{113},
  157603\relax
\mciteBstWouldAddEndPuncttrue
\mciteSetBstMidEndSepPunct{\mcitedefaultmidpunct}
{\mcitedefaultendpunct}{\mcitedefaultseppunct}\relax
\EndOfBibitem
\bibitem[G{\"o}rller-Walrand and Binnemans(1996)G{\"o}rller-Walrand, and
  Binnemans]{GorllerWalrand:1996bt}
G{\"o}rller-Walrand,~C.; Binnemans,~K. \emph{Handbook on the Physics and
  Chemistry of Rare Earths}; Elsevier: Amsterdam, 1996; pp 101--264\relax
\mciteBstWouldAddEndPuncttrue
\mciteSetBstMidEndSepPunct{\mcitedefaultmidpunct}
{\mcitedefaultendpunct}{\mcitedefaultseppunct}\relax
\EndOfBibitem
\bibitem[Kunkel \latin{et~al.}(2016)Kunkel, Bartholomew, Binet, Ikesue, and
  Goldner]{Kunkel:2016fh}
Kunkel,~N.; Bartholomew,~J.; Binet,~L.; Ikesue,~A.; Goldner,~P.
  {High-Resolution Optical Line Width Measurements as a Material
  Characterization Tool}. \emph{J. Phys. Chem. C} \textbf{2016}, \emph{120},
  13725--13731\relax
\mciteBstWouldAddEndPuncttrue
\mciteSetBstMidEndSepPunct{\mcitedefaultmidpunct}
{\mcitedefaultendpunct}{\mcitedefaultseppunct}\relax
\EndOfBibitem
\bibitem[Robertson(2004)]{Robertson:2004fa}
Robertson,~J. {High dielectric constant oxides}. \emph{Eur. Phys. J. Appl.
  Phys.} \textbf{2004}, \emph{28}, 265--291\relax
\mciteBstWouldAddEndPuncttrue
\mciteSetBstMidEndSepPunct{\mcitedefaultmidpunct}
{\mcitedefaultendpunct}{\mcitedefaultseppunct}\relax
\EndOfBibitem
\bibitem[Walther \latin{et~al.}(2015)Walther, Rippe, Yan, Karlsson, Serrano,
  Nilsson, Bengtsson, and Kr{\"o}ll]{Walther:2015hh}
Walther,~A.; Rippe,~L.; Yan,~Y.; Karlsson,~J.; Serrano,~D.; Nilsson,~A.~N.;
  Bengtsson,~S.; Kr{\"o}ll,~S. {High-fidelity readout scheme for rare-earth
  solid-state quantum computing}. \emph{Phys. Rev. A} \textbf{2015}, \emph{92},
  022319\relax
\mciteBstWouldAddEndPuncttrue
\mciteSetBstMidEndSepPunct{\mcitedefaultmidpunct}
{\mcitedefaultendpunct}{\mcitedefaultseppunct}\relax
\EndOfBibitem
\bibitem[Ruggiero \latin{et~al.}(2009)Ruggiero, Le~Gou{\"e}t, Simon, and
  Chaneli{\`e}re]{Ruggiero:2009uu}
Ruggiero,~J.; Le~Gou{\"e}t,~J.-L.; Simon,~C.; Chaneli{\`e}re,~T. {Why the
  two-pulse photon echo is not a good quantum memory protocol}. \emph{Phys.
  Rev. A} \textbf{2009}, \emph{79}, 053851\relax
\mciteBstWouldAddEndPuncttrue
\mciteSetBstMidEndSepPunct{\mcitedefaultmidpunct}
{\mcitedefaultendpunct}{\mcitedefaultseppunct}\relax
\EndOfBibitem
\bibitem[Zhong \latin{et~al.}(2017)Zhong, Kindem, Bartholomew, Rochman,
  Craiciu, Miyazono, Bettinelli, Cavalli, Verma, Nam, Marsili, Shaw, Beyer, and
  Faraon]{Zhong:2017fe}
Zhong,~T.; Kindem,~J.~M.; Bartholomew,~J.~G.; Rochman,~J.; Craiciu,~I.;
  Miyazono,~E.; Bettinelli,~M.; Cavalli,~E.; Verma,~V.; Nam,~S.~W.;
  Marsili,~F.; Shaw,~M.~D.; Beyer,~A.~D.; Faraon,~A. {Nanophotonic rare-earth
  quantum memory with optically controlled retrieval}. \emph{Science}
  \textbf{2017}, \emph{357}, 1392--1395\relax
\mciteBstWouldAddEndPuncttrue
\mciteSetBstMidEndSepPunct{\mcitedefaultmidpunct}
{\mcitedefaultendpunct}{\mcitedefaultseppunct}\relax
\EndOfBibitem
\bibitem[Julsgaard \latin{et~al.}(2013)Julsgaard, Grezes, Bertet, and
  M{\o}lmer]{Julsgaard:2013br}
Julsgaard,~B.; Grezes,~C.; Bertet,~P.; M{\o}lmer,~K. {Quantum Memory for
  Microwave Photons in an Inhomogeneously Broadened Spin Ensemble}. \emph{Phys.
  Rev. Lett.} \textbf{2013}, \emph{110}, 250503\relax
\mciteBstWouldAddEndPuncttrue
\mciteSetBstMidEndSepPunct{\mcitedefaultmidpunct}
{\mcitedefaultendpunct}{\mcitedefaultseppunct}\relax
\EndOfBibitem
\bibitem[Wolfowicz \latin{et~al.}(2014)Wolfowicz, Urdampilleta, Thewalt,
  Riemann, Abrosimov, Becker, Pohl, and Morton]{Wolfowicz:2014ji}
Wolfowicz,~G.; Urdampilleta,~M.; Thewalt,~M. L.~W.; Riemann,~H.;
  Abrosimov,~N.~V.; Becker,~P.; Pohl,~H.-J.; Morton,~J. J.~L. {Conditional
  Control of Donor Nuclear Spins in Silicon Using Stark Shifts}. \textbf{2014},
  \emph{113}, 157601\relax
\mciteBstWouldAddEndPuncttrue
\mciteSetBstMidEndSepPunct{\mcitedefaultmidpunct}
{\mcitedefaultendpunct}{\mcitedefaultseppunct}\relax
\EndOfBibitem
\end{mcitethebibliography}

\end{document}